# CFW: A Collaborative Filtering System Using Posteriors Over Weights of Evidence

**Carl M. Kadie, Christopher Meek, and David Heckerman**
Microsoft Research
Redmond, WA 98052-6399
{kadie,meek,heckerma}@microsoft.com

## Abstract

We describe CFW, a computationally efficient algorithm for collaborative filtering that uses posteriors over weights of evidence. In experiments on real data, we show that this method predicts as well or better than other methods in situations where the size of the user query is small. The new approach works particularly well when the user's query contains low frequency (unpopular) items. The approach complements that of dependency networks which perform well when the size of the query is large. Also in this paper, we argue that the use of posteriors over weights of evidence is a natural way to recommend similar items—a task that is somewhat different from the usual collaborative-filtering task.

## 1 INTRODUCTION

Collaborative filtering (CF) is the task of predicting user preferences over items such as books, television shows, movies, and web pages. CF takes as input a set of items preferred by a particular user as well as sets of preferred items for a collection of users, and returns a list of additional items that the given user will likely prefer (Resnick, Iacovou, Suchak, Bergstorm, & Riedl, 1994; Breese, Heckerman, & Kadie, 1998). For example, if a user queries a CF system for television shows by telling it that she likes "Murder She Wrote" and "Diagnosis Murder", the system may predict that she would also like "60 Minutes" and "Matlock". Some systems generate predictions by first finding other users in the collection who like the same shows as the user, and then suggesting some of the other shows that those users like. Alternatively, some systems use the data to construct a model of user preference. This model is then used to reply to queries. Breese, *et al.* (1998) describe several CF scenarios including binary versus non-binary preferences and implicit versus explicit voting. In the television domain, for example, an implicit voting system using binary preferences would predict which shows you would like based on observations of what shows you watch for (say) five minutes or more. Implicit voting of binary preferences is also important for e-commence. We concentrate on this scenario in this paper.

The creation of the CFW algorithm was motivated by a desire to find a practical prediction algorithm that would work well on small queries and on queries containing low frequency items. By practical, we mean that the method's learning time is not prohibitively long or memory intensive, and those queries can be answered quickly. By low frequency items, we mean those items that few people in the collection have expressed interest. As Zipf's law suggests (Zipf, 1949), low frequency items are common in many domains. Moreover, low frequency items may indicate specialty interests of the user that may prove useful for recommendation. Small queries are of interest because they can be based on the user's immediate context. In the television domain, for example, a query might be one based on the last few shows the you have watched, or perhaps just the show you are currently watching. As you watch the show "Skeleton Warrior", a show relatively few people watch, you might appreciate a recommendation for "WildC.A.T.S.".

In section 2, we describe the CFW algorithm. The algorithm uses numerical methods and Bayes' rule to produce posterior distributions for weights of evidence between all pairs of items in the domain. At query time, it uses these distributions to find the items most likely to be preferred. In section 3, CFW is put to the test against a variety of CF systems for three real domains. In the experiments, CFW performs well for queries of length five and less. In section 4, we look at another use for posteriors over weights of evidence: the determination of item similarity. In addition, we discuss related and future work.

## 2 ALGORITHM

As mentioned, the goal of CF is to predict items that a user likes based on other items liked by the user. The CFW algorithm learns a two-part probabilistic model for this task. Part one is the (marginal) probability that an item is liked for every item in the domain. Part two is a measure of association between every pair of items. As our CF task applies to an (assumed infinite) population of users, these quantities can be thought of as long-run fractions and quantities derived there from. We use a Bayesian approach to reason about the uncertainty in these quantities.

### 2.1 ITEM MARGINALS

For each item $x$, let $\theta_x$ denote the long-run fraction of users who like $x$. CFW assumes that, *a priori*, $\theta_x$ is distributed Beta$(1/2, 1/2)$. Thus, *a posteriori*,

$$\theta_x \sim \text{Beta}\left(a_x + \frac{1}{2}, b_x + \frac{1}{2}\right) \quad (1)$$

where $a_x$ is the number of people in the collection who like item $x$, and $b$ is the number of people in the collection who do not. We use the median value when we need point estimates for the marginal distribution. On data not used in our evaluation, we experimented with more complicated approaches, but none yielded predictions more accurate than this one. For example, we examined (1) other priors such as a Zipf-like model $\theta_x \sim \text{Beta}(\frac{1}{shape} + a, 1 + b)$, where *shape* is a parameter that characterizes the "Zipfiness" of the distribution, (2) point estimates other than median (e.g., various percentiles), and (3) carrying the full distribution through the ranking process.

### 2.2 ITEM ASSOCIATIONS

The measure of item-item association that we use is the weight of evidence.

#### 2.2.1 Weight of Evidence

The weight of evidence for a hypothesis given some event is, roughly speaking, the degree to which observing the event increases the probability of the hypothesis. To be more precise, we define the *weight of evidence from a given item $x_e$ to item $x_h$* as

$$\omega(e:h) = \ln \frac{\theta_{e|h}}{\theta_{e|\bar{h}}} \quad (2)$$

where $\theta_{e|h}$ and $\theta_{e|\bar{h}}$ are the long-run fractions of people who like $x_e$ among those people who like $x_h$ and who do not like $x_h$, respectively. By Bayes' rule, we have

$$\ln \frac{\theta_{h|e}}{1 - \theta_{h|e}} = \omega(e:h) + \ln \frac{\theta_h}{1 - \theta_h} \quad (3)$$

where $\theta_{h|e}$ is the long-run fraction of people who like $x_h$ among those people who like $x_e$.

#### 2.2.2 Inferring the Weight of Evidence

We can infer or estimate a weight of evidence $\omega(e:h)$ given the known preferences for $x_e$ and $x_h$ in the user collection. This data can be expressed as counts in a $2\times 2$ contingency table: 

| $a$ | $b$ |
|---|---|
| $c$ | $d$ |

where $a$ is the number of people in the collection who like neither item, $b$ is the number who like $x_e$ but not $x_h$, $c$ is the number who like $x_h$ but not $x_e$, and $d$ is the number who like both items. For convenience we will write this table in flattened form: $C = [[a,b],[c,d]]$.

If we use a maximum-likelihood approach to estimate both $\theta_{e|h}$ and $\theta_{e|\bar{h}}$, then we obtain the estimate

$$\hat{\omega}(e:h) = \ln\left(\frac{\frac{d}{c+d}}{\frac{b}{a+b}}\right) \quad (4)$$

When $b = 0$ or $c + d = 0$, however, the results are undefined. Furthermore, for small counts, the results are unstable. As mentioned, we use instead use a Bayesian approach. (See Section 4.1.2 for an alternative approach.)

In this approach, we assume that $\theta_e$ and $\theta_h$ are known, and encode our uncertainty about $\omega(e:h)$ using the prior distribution $p(\omega(e:h)|\theta_e, \theta_h)$. We then infer a posterior distribution for $\omega(e:h)$ given $C = [[a,b],[c,d]]$ using Bayes' rule:

$$p(\omega(e:h)|C, \theta_h, \theta_e) = \\ \alpha \cdot p(C|\omega(e:h), \theta_h, \theta_e) \cdot p(\omega(e:h)|\theta_h, \theta_e) \quad (5)$$

where $\alpha$ is a normalization constant. The likelihood term is given by

$$p(C|\omega(e:h), \theta_h, p(x_e)) = \binom{N}{a,b,c,d} \theta_{\bar{e}\bar{h}}^a \theta_{e\bar{h}}^b \theta_{\bar{e}h}^c \theta_{eh}^d$$

where $\theta_{eh}$ is the long-run fraction of people who like both $x_e$ and $x_h$, and so on. Note that each long-run fraction in this expression can be computed from $\omega(e:h)$, $\theta_e$, and $\theta_h$ using the rules of probability.

In our system, we make the assumption that all weights of evidence are mutually independent, and separately compute a posterior for each weight of evidence. This assumption will almost always be inconsistent. For example, if we know $\omega(x:y)$ and $\omega(y:z)$, then the possible values of $\omega(x:z)$ are constrained. Nonetheless, we make this assumption because it leads to a computationally efficient approach that predicts well in practice. In Section 2.3, we discuss a consistent albeit less tractable approach.

Experimenting with data not used in our formal evaluation, we found prediction accuracy to be rather insensitive to (smooth) priors on $\omega(e:h)$. In our experiments, we use

$$\omega(e:h) \sim \text{Uniform}(-6, 6)$$

for all $x_e$ and $x_h$. We note that, for some values of $\theta_e$ and $\theta_h$, some values for $\omega(e:h)$ at the extremes of this range are impossible. In those situations, we use a numerical binary search to find the allowed subrange and take as the prior the uniform distribution over that subrange.

We approximate posteriors for weights of evidence using the following two-pass numerical approximation technique:

- Split the domain of $p(\omega)$ into fifteen equal segments. Because the prior is uniform and the segments have identical widths, the segments have equal priors.

- Using point values at the center of each segment, find the discrete posterior distribution of the segments.

- Find the segment that precedes the first segment with posterior probability greater than 0.00001. If no such segment exists, choose the first segment. In addition, find the segment that follows the last segment with a posterior probability greater than 0.00001. If no such segment exists, choose the last segment.

- Combine the identified segments into one segment and split that segment into fifteen equal segments.

- We now have the leftover old segments of one width and new segments with a second width. Assign each segment a prior probability proportional to its width.

- Using point values at the center of each segment, find the discrete posterior distribution of the segments.

- Return this distribution.

Figure 1 shows the approximated posterior for two sets of counts using our approach.

Given a posterior over $\omega(e:h)$, we determine a point estimate for it. As we shall see, this point estimate will increase the computational efficiency of our approach. The point estimate we use is motivated by the case where there is only one item of evidence available when making recommendations. In this situation, under our previous assumption that $\omega(e:h)$ for each $h$ are mutually independent, the list of recommended items should be sorted by the expected posterior of each item:

$$\langle \theta_{h|e} \rangle = \int \theta_{h|e}\ p(\omega|C)\ d\omega \qquad (6)$$

where $\omega$ and $p(\omega|C)$ are abbreviations for $\omega(e:h)$ and $p(\omega(e:h)|C, \theta_h, \theta_e)$, respectively. This sort order will maximize the user's expected utility as defined in Section 3. Inverting Equation 3 and using $\omega_0 = ln(\theta_h/(1-\theta_h))$, we rewrite Equation 6 as

$$\langle \theta_{h|e} \rangle = \int \frac{e^{\omega+\omega_0}}{1+e^{\omega+\omega_0}}\ p(\omega|C)\ d\omega \qquad (7)$$

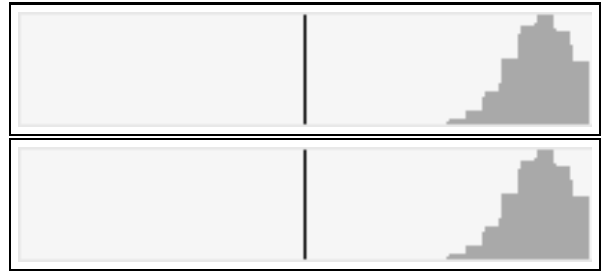

Figure 1: Posterior over $\omega(e:h)$ for [[1621,3],[10,4]] and [[1624,3],[10,1]]. The scale is from -6.0 to 6.0. The vertical line marks 0.0.

Given our discrete estimate for $p(\omega)$, we approximate this integral as

$$\langle \theta_{h|e} \rangle \simeq \sum_i \frac{e^{\omega^i+\omega_0}}{1+e^{\omega^i+\omega_0}}\ p(\omega^i|C) \qquad (8)$$

where $\omega^i$ is the value of $\omega$ at the center of the $i$th segment. Finally, we substitute $\langle \theta_{h|e} \rangle$ into Equation 3 to obtain $\omega_{\text{eff}}(e:h)$, our point estimate for $\omega(e:h)$:

$$\omega_{\text{eff}}(e:h) = \ln \frac{\langle \theta_{h|e} \rangle}{1 - \langle \theta_{h|e} \rangle} - \omega_0 \qquad (9)$$

We call $\omega_{\text{eff}}(e:h)$ the *effective weight of evidence from a given item $x_e$ to item $x_h$*.

### 2.2.3 Computational Issues

To compute the posterior over $\omega(e:h)$ for each pair of items, our system must determine how many users like each pair of objects (count $d$). The biggest problem in calculating these counts is the $n^2$ memory that it could take. When necessary, we avoid this problem by partitioning all items into $m$ subsets of roughly equal size and passing over the data $m$ times. On each pass through the data, we determine counts $d$ for all items $x_e$ in the $m$th partition and all items $x_h$ in the whole domain.

In order to make predictions quickly at query time, we do not save all pair wise associations. Instead, for each $x_e$, we store a relatively small number ($f = 40$) of associated items $x_h$. The items we choose depend on how the lists are to be used. For our evaluation of recommendation accuracy (Section 3), where our goal is to find items that are likely to be preferred given the query, we cache the $f$ $x_h$'s with the highest point estimates for $\theta_{h|e}$. For the task of finding similar items (i.e., items $x_h$ where $\omega(e:h)$ is large), we cache the $f$ $x_h$'s that have the largest point estimates for the weight of evidence $\omega(e:h)$.

Our system uses two techniques to speed up processing. First, it caches posterior calculations, remembering the posteriors for the first 100,000 distinct vectors $[[a,b],[c,d]]$. Second, it approximately bounds each posterior using a fast computation based on Equation 4 in which counts are varied by $\pm 1$. If no value for $\omega(e:h)$ meets the criteria for making it on the list, then $x_h$ is skipped.

### 2.3 APPLYING THE MODEL

At query time, CFW produces recommendations in a straightforward manner. In particular, CFW takes each $x_e$ in the user's query and looks up its list of associated $x_h$s. Then, for each $x_h$ found, it combines the point estimate for $\theta_h$ (its posterior median), and the point estimates for the weights of evidence (the effective weights of evidence) associated with $x_h$ to determine a point estimate for $\theta_{h|e}$. The recommendation list is then sorted by this posterior estimate.

When $x_h$ is on the list of more than one item from the user's query, we need to combine the evidence for $x_h$. In one approach, an approximation to Naive Bayes, we sum the effective weights of evidence. In a more conservative approach, we use the maximum of the effective weights. We evaluate both approaches, denoted CFWp and CFWm, respectively.

Using these simple methods ignores what we know about the pair wise relationship between pieces of evidence. One (consistent) method for using all pair wise information would be to estimate a log-linear model with all pair wise interactions and use this model to calculate the required conditional probabilities (e.g., Agresti, 1990). To remain (somewhat) tractable, a log-linear model should be learned for each hypothesis and the given query at query time. Obviously, the run time of this approach would be significantly slower than the simpler approaches.

### 3 EMPIRICAL ANALYSIS

We evaluated CFW (CFWp and CFWm) on the three datasets used by Heckerman, Chickering, Meek, Rounthwaite, & Kadie (2000). The datasets are (1) Nielsen, data about whether or not a user watched five or more minutes of network TV shows aired during a two-week period in 1995 (made available by Nielsen Media Research), (2) MS.COM, which records whether or not anonymous users of microsoft.com on one day in 1996 visited areas ("vroots") of the site (available on the Irvine Data Mining Repository), and (3) MSNBC, which records whether or not visitor to MSNBC on one day in 1998 read stories among the most popular 1001 stories on the site. Basic statistics for these datasets are given in Table 1.

Table 1: Number of users, items, and average items per user (in training set) for the datasets used in our evaluation.

|  | Dataset | | |
| --- | --- | --- | --- |
|  | MS.COM | Nielsen | MSNBC |
| Users in train | 32711 | 1638 | 10000 |
| Users in test | 5000 | 1637 | 10000 |
| Total items | 294 | 203 | 1001 |
| Items per user | 3.02 | 8.64 | 2.67 |

Following Heckerman *et al.* (2000), our criterion attempts to measure a user's expected utility for a list of recommendations. Of course, different users will have different utility functions. The measure we use provides what we believe to be a good approximation across many users.

The scenario we imagine is one where a user is shown a ranked list of items and then scans that list for preferred items starting from the top. At some point, the user will stop looking at more items. Let $p(k)$ denote the probability that a user will examine the $k$th item on a recommendation list before stopping his or her scan, where the first position is given by $k = 0$. Then, a reasonable criterion is

$$\text{cfaccuracy}_1(\text{list}) = \sum_k p(k)\ \delta_k,$$

where $\delta_k$ is 1 if the item at position $k$ is preferred and 0 otherwise. To make this measure concrete, we assume that $p(k)$ is an exponentially decaying function:

$$p(k) = 2^{-k/a}, \qquad (10)$$

where $a$ is the "half-life" position—the position at which an item will be seen with probability 0.5. In our experiments, we use $a = 5$.

In one possible implementation of this approach, we could show recommendations to a series of users and ask them to rate them as "preferred" or "not preferred". We could then use the average of cfaccuarcy$_1$(list) over all users as our criterion. Because this method is extremely costly, we instead use an approach that uses only the data we have. In particular, as already described, we randomly partition a dataset into a training set and a test set. Each case in the test set is then processed as follows. First, we randomly partition the user's preferred items into *input* and *measurement* sets. The input set becomes the query to the CF model, which in turn outputs a list of recommendations. Finally, we compute our criterion as

$$\text{cfaccuracy(list)} = \frac{100}{N} \sum_{i=1}^{N} \frac{\sum_{k=0}^{R_i-1} \delta_{ik} \ p(k)}{\sum_{k=0}^{M_i-1} p(k)}, \quad (11)$$

where $N$ is the number of users in the test set, $R_i$ is the number of items on the recommendation list for user $i$, $M_i$ is the number of preferred items in the measurement set for user $i$, and $\delta_{ik}$ is 1 if the $k$th item in the recommendation list for user $i$ is preferred in the measurement set and 0 otherwise. The denominator in Equation 11 is a per-user normalization factor. It is the utility of a list where all preferred items are at the top. This normalization allows us to more sensibly combine scores across measurement sets of different size.

We looked at five protocols. In the all-but-1 protocol, the query set contained all the items except one. In the given-$k$ protocols, the query sets contained exactly $k$ items. As in Breese *et al.* (1998) and Heckerman *et al.* (2000), we used $k = \{2, 5, 10\}$. To this, we added $k = 1$.

In addition to CFW, the CF systems tested included CR+, the best performing memory-based method reported in Breese *et al.*(1998). (Microsoft Site Server v3.0, Commerce Edition uses this CF algorithm.) From Heckerman *et al.*(2000), we used Bayesian-network learner (BN) and the Dependency-network learner (DN). Of these only DN is practical in that, unlike BN, it learns in a reasonable amount of time and, unlike CR+, it can evaluate queries quickly. (Microsoft Commerce Server 2000 and 2002 use DN.)

We also show the score of a recommendation list made up of all items in popularity order and label these results *baseline*.

### 3.1 PREDICTION ACCURACY

Table 2 shows the accuracy of the five methods across three datasets and five protocols. A score in boldface is either the winner or statically tied with the winner.

We use ANOVA with $\alpha = 0.1$ to test for statistical significant (McClave & Dietrich, 1988). When the difference between two scores in the same column exceed the value of RD (required difference), the difference is significant.

The results show CFW doing very well compared to other methods when the size of the query is small, with CFWm generally outperforming CFWp (Naive Bayes). Specifically, in "Given 1", although CFWm loses to CR+, CR+'s relative slowness at evaluating queries makes it impractical. (On "Given 1", CFWm and CFMp are the same.) Excluding CR+, CFWm has the top score on "Given 1" in all the domains. On "Given 2" and "Given 5", CFWm either wins or is in a statistical tie with the winner in all the domains. On "Given 10" and "All But 1", CFWm loses in two of the six experiments. Contrast this with DN, the other practical method, which loses only once. Thus, CFWm and DN may be complimentary, and can be combined in a simple manner based on query length.

### 3.2 SHORT QUERIES AND LOW FREQUENCY ITEMS

As we have seen, CFW performs better on short queries. A reasonable explanation for this observation is that the validity of the assumptions underlying CFW (the naive Bayes assumptions for CFWp and similarly restrictive assumptions for CFWm) degrade as the length of the query increases.

Another observation is that CFW performs better than DN on short queries. We hypothesize that this is the case because CFW uses every item in the user's query, even low frequency items. DN, in contrast, uses decision trees, which in our experience often leave out low frequency items in favor of higher frequency items that contain the same information. To test this hypothesis, we performed experiments with short queries ("Given 1" and "Given 2") in which all items were either low frequency (liked by fewer users than the median item) or high frequency (liked by more users than the median item). The results are shown in Table 3. As we would expect from the previous results, CFWm does better than DN on all these small queries. In all cases, moreover, CFWm's relative performance is better on queries containing low frequency items compared to queries containing high frequency items. This confirms that CFWm's performance is due at least in part to its better exploitation of low frequency items.

### 3.3 PREDICTION RUN TIMES

Table 4 shows total number of queries processed per second for each of the experimental conditions on a 2.2 GHz Pentium 4 running Windows XP Pro. To the resolution of timing (1 second), CFW is the same as DN on MS.COM and Nielsen, but slower on MSNBC. Both CFW and DN are substantially faster than CR+ for both MS.COM and Nielsen.

Table 2: CF accuracy for the MS.COM, Nielsen, and MSNBC datasets. Higher scores indicate better performance. The winning score and scores not statistically significantly different than the winner are shown in boldface.

|           | MS.COM  |         |         |          |           |
|-----------|---------|---------|---------|----------|-----------|
| Algorithm | Given 1 | Given 2 | Given 5 | Given 10 | All But 1 |
| BN        | 54.96   | **53.18** | **52.48** | **51.64** | **66.54** |
| DN        | 53.67   | 52.68   | **52.54** | **51.48** | **66.60** |
| CFWm      | 55.58   | **53.59** | **52.20** | 49.85    | 63.29     |
| CFWp      | 55.58   | **53.67** | 47.79   | 42.12    | 62.00     |
| CR+       | **56.48** | **53.42** | 49.64   | **47.74** | 63.59     |
| $\overline{RD}$ | 0.67 | 0.97 | 2.24 | 6.13 | 1.09 |
| Baseline  | 47.19   | 43.37   | 39.34   | 39.32    | 49.77     |
|           | Nielsen |         |         |          |           |
| Algorithm | Given 1 | Given 2 | Given 5 | Given 10 | All But 1 |
| BN        | 21.37   | 24.99   | **30.03** | **33.84** | **45.55** |
| DN        | 20.92   | 24.20   | **29.71** | **33.80** | **44.30** |
| CFWm      | **23.53** | **27.24** | **30.84** | **33.41** | **43.78** |
| CFWp      | **23.53** | **26.77** | 28.32   | 29.00    | 39.02     |
| CR+       | **23.50** | **27.46** | 29.49   | 29.76    | 39.74     |
| $\overline{RD}$ | 1.15 | 1.21 | 1.59 | 2.31 | 2.37 |
| Baseline  | 12.89   | 12.65   | 12.72   | 12.92    | 13.59     |
|           | MSNBC   |         |         |          |           |
| Algorithm | Given 1 | Given 2 | Given 5 | Given 10 | All But 1 |
| BN        | 40.96   | 40.34   | **34.20** | **30.39** | **49.58** |
| DN        | 39.52   | 38.84   | **32.53** | **30.03** | 48.05     |
| CFWm      | **43.96** | **42.20** | **33.90** | 25.94    | **49.35** |
| CFWp      | **43.96** | **41.91** | **32.39** | **29.41** | 48.66     |
| CR+       | **44.51** | **42.74** | **33.49** | **29.10** | **49.94** |
| $\overline{RD}$ | 0.63 | 0.93 | 1.88 | 4.34 | 0.99 |
| Baseline  | 32.27   | 28.73   | 20.58   | 14.93    | 32.94     |

Table 3: The effect of small queries containing either all low frequency item or all high frequency items. All results except the 4th line are statistically significant ($\alpha = 0.10$).

| Domain  | Protocol | Range      | DN    | CFWm  | Diff. |
|---------|----------|------------|-------|-------|-------|
| MS.COM  | Given 1  | Low Freq.  | 32.83 | 44.32 | 11.48 |
|         |          | High Freq. | 53.37 | 55.06 | 1.68  |
|         | Given 2  | Low Freq.  | 29.16 | 40.62 | 11.46 |
|         |          | High Freq. | 52.11 | 52.74 | 0.63  |
| Nielsen | Given 1  | Low Freq.  | 16.71 | 20.00 | 3.29  |
|         |          | High Freq. | 19.93 | 22.33 | 2.40  |
|         | Given 2  | Low Freq.  | 16.72 | 20.48 | 3.76  |
|         |          | High Freq. | 22.51 | 24.46 | 1.95  |
| MSNBC   | Given 1  | Low Freq.  | 20.57 | 28.52 | 7.96  |
|         |          | High Freq. | 39.45 | 43.37 | 3.92  |
|         | Given 2  | Low Freq.  | 16.78 | 23.64 | 6.86  |
|         |          | High Freq. | 38.17 | 41.12 | 2.95  |

## 4 DISCUSSION

In this section we look at extensions, related work, and future research.

### 4.1 EXTENSIONS

We have implemented two extensions to CFW. The first allows it to limit recommendations to items having a strong association (with high probability) to items in the query. The second uses a simpler method to calculate a point estimate for the weight of evidence.

#### 4.1.1 Recommending Similar Items

So far, we have concentrated on the prediction of likely items. In some situations, however, it may be more desirable to recommend similar items. The weight of evidence is a natural measure of similarity; and our CFW algorithm can be easily adapted to provide such recommendations.

As an example, consider the single-item query "Matlock" in the Nielsen domain. Table 5 lists other TV shows having the highest values of posterior probability. Note that, although all three shows have a high posterior probability, only "Commish" and "Murder She Wrote" much higher weight of evidence than "ABC World News". This is consistent with our understanding that these two shows are more similar to "Matlock" than is "ABC World News".

Table 5: Recommendation weights given the query "Matlock".

| Item             | $\theta_{h_e}$ | $\omega(e:h)$ |
|------------------|----------------|---------------|
| Commish          | 0.49           | 2.6           |
| ABC World News   | 0.43           | 1.2           |
| Murder She Wrote | 0.40           | 1.9           |

Table 4: Number of queries processed per second. Larger is better.

| MS.COM | | | | | |
|---|---|---|---|---|---|
| | Given 1 | Given 2 | Given 5 | Given 10 | All But 1 |
| BN | 22 | 22 | 21 | 11 | 22 |
| DN | 105 | 105 | 73 | 25 | 105 |
| CFWm | 105 | 96 | 73 | 20 | 96 |
| CFWp | 105 | 101 | 73 | 25 | 99 |
| CR+ | 22 | 18 | 12 | 7 | 14 |

| Nielsen | | | | | |
|---|---|---|---|---|---|
| | Given 1 | Given 2 | Given 5 | Given 10 | All But 1 |
| BN | 86 | 87 | 84 | 80 | 91 |
| DN | 146 | 145 | 131 | 120 | 162 |
| CFWm | 122 | 130 | 115 | 80 | 112 |
| CFWp | 133 | 118 | 115 | 80 | 112 |
| CR+ | 91 | 93 | 77 | 60 | 77 |

| MSNBC | | | | | |
|---|---|---|---|---|---|
| | Given 1 | Given 2 | Given 5 | Given 10 | All But 1 |
| BN | 33 | 33 | 30 | 21 | 33 |
| DN | 47 | 53 | 49 | 30 | 55 |
| CFWm | 32 | 32 | 29 | 19 | 32 |
| CFWp | 32 | 32 | 28 | 19 | 32 |
| CR+ | 53 | 43 | 31 | 18 | 39 |

As we have mentioned, the effective weight of evidence serves us well when predicting likely items. In contrast, when the task is identifying similar items, we have found it useful to use a point estimate at a conservative percentile (typically 5%). That is, we order associations by the $\omega(e:h)$ that we are 95% sure of achieving. For example, for the first distribution in Figure 1, we use a value for $\omega(e:h)$ of 3.8, rather than its effective value of 4.9. This conservative strategy down weights coincidental associations while still extracting information from low frequency items and low count associations.

Table 6 illustrates the effect of this conservative strategy in the Nielsen domain given the query of "Matlock". The first and second recommendation lists come from the conservative (5%) and effective $\omega(e:h)$ strategies, respectively. Subjectively, the first list contains more similar items near the top.

#### 4.1.2 Simple Estimation of $\omega(e:h)$

A simple approach for estimating the weight of evidence is to use Equation 2 and estimate $\theta_{e|h}$ and $\theta_{e|\bar{h}}$ separately. To perform comparisons with the approach described in Section 2.2.2, we implemented this simple approach using $(1/2 + a)/(1 + a + b)$ and $(1/2 + c)/(1 + c + d)$ as the estimates for $\theta_{e|\bar{h}}$ and $\theta_{e|h}$, respectively.

For the task of predicting likely items, this simple approach was competitive with that of CFW. In contrast, when used to generate similar items, the simple approach appeared to introduce "noise" into the recommendation list. This effect is illustrated in part (c) of Table 6. In this list, some shows such as "Pizza Hut Col Bsktbl-SU" do not appear to be similar to the query "Matlock".

### 4.2 RELATED WORK

Classical measures (estimates) of association for 2x2 tables have been studied for decades. Among the most popular is the cross-product ratio (also called the odds ratio), which on table $[[a, b], [c, d]]$ is $\hat{\theta} = ad/bc$. Other measures include (1) the log of the cross-product ratio, (2) Yule's Q, which is the 2x2 version of the Gamma measure, (3) Phi, which is the 2x2 version of Pearson's correlation coefficient, and (4) various "tau" measures (Reynolds, 1984; Gibbons, 1993).

The accuracy of these measures of association can be characterized with the asymptotic standard error (ASE) of the estimator. For example, the standard error for the log of the cross-product ratio (Agresti, 1990) is given by

$$se(\log \hat{\theta}) = (1/a + 1/b + 1/c + 1/d)^{1/2} \qquad (12)$$

The standard error can then be used to estimate a confidence interval due to the asymptotic normality of the estimator $\log \hat{\theta}$. Although we prefer our Bayesian approach, these measures are viable alternatives.

Recent work by Sarwar, Karypis, Konstan, & Riedl (2001) on a different collaborative filtering scenario has also looked at creating fast algorithms by learning item-to-item tables. In their scenario, the ratings are real valued and given explicitly by users. Their goal is to predict the numeric rating that a user would assign to a held-out item. Their prediction method uses measures of similarity based on cosine similarity and Pearson-$r$ correlation. In contrast, in our scenario, ratings are binary and implicit, and our goal is to create a recommendation list with high utility. This goal motivates our use of weight of evidence and posterior probability. See Breese et al. (1998) for a detailed discussion of these two collaborative filtering scenarios.

## 4.3 SUMMARY AND FUTURE WORK

The CFW algorithm is a computationally efficient method for collaborative filtering. The algorithm starts by estimating the margins for all items in the domain and the posterior weight of evidence for all pairs of items in the domain. Because the approach makes use of a posterior on weight of evidence, it can extract information from low count cases that might otherwise either be ignored or give unstable statistics. This ability, in turn, allows CFW to find interesting relationships even among low frequency items. Low frequency items are of interest both because of their prevalence in domains with Zipf-like distributions and because they may be the valuable for understanding a user's special interest.

We have considered two approaches for combining weights of evidence: plus (Naive Bayes) and max. Neither alternative exploits the information available about the two-way associations between query items. As discussed in Section 2.3, a computationally more expensive approach exists that consistently use this information. An interesting area for future work would be to evaluate the benefits of this alternative.

Our main approach uses a numerical procedure for approximating the posterior distribution for weight of evidence. It would be useful to compare alternative approaches for approximating the posterior with respect to the cost of computing the approximation and the sensitivity of the prediction results. In addition, preliminary experiments have demonstrated that simplified procedures for computing point estimates for the weights of evidence perform competitively in some scenarios. Further investigation is needed to better understand the scenarios in which the simplified approaches are appropriate. Further investigations are also needed to understand the sensitivity of the results on the form of prior for the margins, and the effect of assuming that unspecified items are missing at random.

Also, our experiments show that, unlike decision-tree-based collaborative filtering methods, CFW makes excellent recommendations when queried on low frequency items. While one could imagine increasing the number of low frequency items in the trees, this would increase the size of the trees substantially. A more realistic alternative would be to use query length to switch between a CFW and DN recommender.

Finally, we described a variation of the CFW algorithm for identifying similar items. We would like to do human subject studies to better understand those situations in which people prefer such recommendations to those based on probability.

## Acknowledgements

We thank Eric Horvitz for his comments on this work. Datasets for this paper were generously provided by Nielsen Media Research (Nielsen), Microsoft Corporation (MS.COM), and Microsoft Corporation and Steve White (MSNBC).

## References

[Agresti, 1990] Agresti, A. (1990). *Categorical Data Analysis*. John Wiley & Sons, New York.

[Breese et al., 1998] Breese, J. S., Heckerman, D., and Kadie, C. (1998). Empirical analysis of predictive algorithms for collaborative filtering. In *Proceedings of Fourteenth Conference on Uncertainty in Artificial Intelligence*, Madison, Wisconsin. Morgan Kaufmann.

[Gibbons, 1993] Gibbons, J. D. (1993). *Nonparametric Measures of Association*. Sage Publications, Newbury Park, CA.

[Heckerman et al., 2000] Heckerman, D., Chickering, D., Meek, C., Rounthwaite, R., and Kadie, C. (2000). Dependency networks for inference, collaborative filtering, and data visualization. *Journal of Machine Learning Research*, 1:49–75.

[McClave and Dieterich, 1988] McClave, J. and Dieterich, F. (1988). *Statistics*. Dellen Publishing Company.

[Resnick et al., 1994] Resnick, P., Iacovou, N., Suchak, M., Bergstorm, P., and Riedl, J. (1994). GroupLens: An Open Architecture for Collaborative Filtering of Netnews. In *Proceedings of ACM 1994 Conference on Computer Supported Cooperative Work*, pages 175–186, Chapel Hill, North Carolina. ACM.

[Reynolds, 1984] Reynolds, H. T. (1984). *Analysis of Nominal Data, 2nd Edition*. Sage Publications, Newbury Park, CA.

[Sarwar et al., 2001] Sarwar, B. M., Karypis, G., Konstan, J. A., and Reidl, J. (2001). Item-based collaborative filtering recommendation algorithms. In *Proceedings of the Tenth International World Wide Web Conference*, pages 285–295.

[Zipf, 1949] Zipf, G. (1949). *Human behavior and the principle of least effort*. Addison–Wesley.

Table 6: Similar items ($\omega(e:h) \geq 1.0$) given the query "Matlock" for (a) CFW at the fifth percentile, (b) CFW using the effective $\omega(e:h)$, and (c) simple estimates for $\omega(e:h)$.

| Item | $\omega(e:h)$ |
|---:|---:|
| Commish | 2.3 |
| Murder She Wrote | 1.6 |
| Diagnosis Murder | 1.5 |
| Day One | 1.3 |
| Nightside M-F-Mon | 1.1 |
| 60 Minutes | 1.0 |

| Item | $\omega(e:h)$ |
|---:|---:|
| Commish | 2.6 |
| Nightside M-F-Mon | 2.2 |
| Murder She Wrote | 1.9 |
| Up To-Minute:Sunday | 1.9 |
| Diagnosis Murder | 1.8 |
| Day One | 1.7 |
| Name Your Adventure | 1.6 |
| *3 shows not shown* | |
| 60 Minutes | 1.4 |
| *20 shows not shown* | |
| CBS Morning News | 1.0 |

| Item | $\omega(e:h)$ |
|---:|---:|
| Commish | 2.6 |
| Nightside M-F-Mon | 2.5 |
| Up To-Minute:Sunday | 2.4 |
| Murder She Wrote | 1.9 |
| Diagnosis Murder | 1.8 |
| Name Your Adventure | 1.7 |
| Day One | 1.7 |
| *... 3 shows not shown ...* | |
| Pizza Hut Coll Bsktbl-SU | 1.5 |
| *... 2 shows not shown ...* | |
| 60 Minutes | 1.3 |
| *... 20 shows not shown ...* | |
| Cybil | 1.0 |